# Compute, Time and Energy Characterization of Encoder-Decoder Networks with Automatic Mixed Precision Training

Siddharth Samsi, Michael Jones, Mark M. Veillette
*MIT Lincoln Laboratory*
240 Wood Street Lexington, MA 02421

*Abstract*—Deep neural networks have shown great success in many diverse fields. The training of these networks can take significant amounts of time, compute and energy. As datasets get larger and models become more complex, the exploration of model architectures becomes prohibitive. In this paper we examine the compute, energy and time costs of training a U-Net based deep neural network for the problem of predicting short term weather forecasts (called precipitation Nowcasting). By leveraging a combination of data distributed and mixed-precision training, we explore the design space for this problem. We also show that larger models with better performance come at a potentially incremental cost if appropriate optimizations are used. We show that it is possible to achieve a significant improvement in training time by leveraging mixed-precision training without sacrificing model performance. Additionally, we find that a 1549% increase in the number of trainable parameters for a network comes at a relatively smaller 63.22% increase in energy usage for a UNet with 4 encoding layers.

## I. INTRODUCTION

Training deep learning models requires significant amounts of computational resources [1], [2]. This is compounded by the fact that the identification of the best model for a given task involves hyperparameter tuning as well as a search for an appropriate model architecture. Thus, optimizing performance of deep neural networks matters - not only for accelerating scientific discovery and insight, but also for reducing the energy costs of training complex models. In this paper we present results of experiments related to training deep neural networks using *mixed precision training* [3]. In mixed precision training, model training operations such as matrix multiplication, convolutions, activations, and gradient computations are carried out in half-precision 16-bit floating point (FP16), while the weight updates are performed on a 32-bit floating point (FP32) copy of the model weights. Using FP16 for mathematical operations has the effect of significantly increasing performance while reducing memory consumption and energy usage.

This material is based upon work supported by the Assistant Secretary of Defense for Research and Engineering under Air Force Contract No. FA8721-05-C-0002 and/or FA8702-15-D-0001. Any opinions, findings, conclusions or recommendations expressed in this material are those of the author(s) and do not necessarily reflect the views of the Assistant Secretary of Defense for Research and Engineering.

Obtaining the full benefit of mixed-precision is not always straightforward. First and foremost, the hardware must support FP16 operations. Second, for best results the model architecture must be well-aligned with hardware specifications. For example, for Tensor Cores on NVIDIA Volta GPUs, maximum performance in matrix multiple is achieved when certain matrix sizes are multiples of 8 [4]. In the context of convolutional neural networks (CNNs), this corresponds to convolutional layers with filter sizes that are multiples of 8. Additionally, larger batch sizes, layers, number of filters can also result in better compute performance due to more efficient memory accesses and reduced overheads.

In this paper, we provide various benchmarks of automatic mixed precision (AMP) training on a standard U-Net model architecture [5]. U-Net is a very common architecture that has been applied to a large number of applications, including image-to-image translation, natural language processing, image segmentation, and others. We use this model for the task of generating short term weather forecasts, called nowcasts. The experiments described in this work use the Storm EVent Imagry (SEVIR) weather dataset[1]. SEVIR events consist of weather image sequences for over 10,000 events that occurred over the US. These events cover 384 km x 384 km patches and span 4 hours. Using this dataset, we train a U-Net to perform the task of predicting $M$ future frames of weather radar imagery given $N$ previous frames. Our experiments demonstrate the speedup of mixed precision training across several variations of the U-Net architecture, as well as a reduction in memory and energy usage. In addition, we show that training using FP16 has little to no impact on the quality of the nowcast model.

## II. PRIOR WORK

Nowcast is the term used for high-resolution, short-term weather forecasts (typically 1-2 hours) of precipitation, cloud coverage or other meteorological quantities [6], [7]. These types of forecasts are of high value in public safety, air traffic control, tactical mission planning and many other areas where high resolution and rapidly updating forecasts are needed. Previous work on deep learning for nowcasting includes convolutional Long Short Term Memory (ConvLSTM) models [8], which showed that DNNs are capable of generating forecasts. Recent work also involves the use of recurrent architectures [9] and fully convolutional networks [7], [10] for precipitation nowcasting.

---

[1]https://registry.opendata.aws/sevir/

The data used for precipitation nowcasting consists of a sequence of images depicting precipitation intensity. This is represented using vertically integrated liquid (VIL) [11] which estimates liquid content in the atmosphere from weather radar measurements. The SEVIR dataset was used to generate training and testing data. We frame nowcasting as a future prediction task where the model input consists of 13 VIL images sampled at 5 minute intervals. The model is trained to produce the next 12 images in the sequence, corresponding to the next hour of weather. Data from SEVIR was first extracted and processed into 44,760 sequences for training and an independent set of 12,144 sequences for testing the fit model. This was done by splitting each SEVIR event into 3 input and output sequences. The model input was of size Nx384x384x13 and the output sequence was Nx384x384x12 pixels, where $N$ is the batch size.

## III. Model Architectures

The nowcast model uses variants of the U-Net architecture [5] with configurable encoders and decoder blocks. The U-Net architecture consists of two phases, an encoder and a decoder. Each of these phases is made of $d$ sequential blocks, which are combinations of layers and activation functions. The number of decoder blocks and filters in each layer are the same as those in the parallel encoder block. Figure 1 shows one such configuration of the encoding and decoding blocks, along with the details of each block. This configuration is used as a baseline model. Table I shows the full set of U-Net configurations explored in this work. The encoder block used here consists of two Conv2D, BatchNorm, Relu sequences, and finalized with a 2x2 MaxPool2d layer. The decoder portion of the network consists of the same number of layers as the encoder blocks. Each decoder block consists of a Conv2DTranspose followed by skip connection with the parallel encoder block, followed by two Conv2D, BatchNorm, Relu sequences. The final layer of the network is Conv2D with linear activation, configured with the appropriate number of outputs depending on the application. The final layer of the network is Conv2D with linear activation, configured with the appropriate number of outputs.

We parameterize different U-Net architectures with two factors: the depth $d$ which controls the number of encoder/decoder blocks, and the starting filter size $f$ which controls the number of filters in the first encoding block. Subsequent encoding (decoding) blocks using double (half) the filters of the previous block. We will use the notion "U$d$-$f$" to describe a given model architecture. The baseline network drawn in figure 1 shows U4-32.

Training time and the quality of model output are largely affected by these hyperparameters, as well as batch sizes used for training, loss function and number of epochs. Table I shows the different U-Net configurations and the resulting number of trainable parameters. As seen in this table, models with four layers can have as many as 273 million unknown parameters simply as a result of increasing the number of filters per layer.

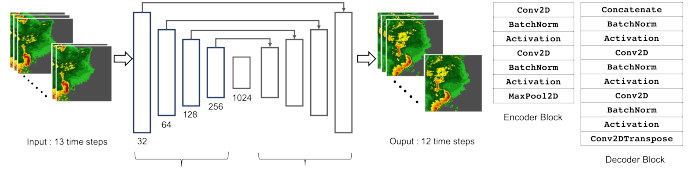

Fig. 1. Example U-Net model architecture used in this paper: This example shows the nowcast workflow where the inputs data consists of an hour of weather radar images in 13 time steps and the output is the predicted weather for the next hour in increments of 5 minutes, resulting in 12 predicted images.

TABLE I
Variations on U-Net architecture explored in this work. The notation "U$d$-$f$" denotes a U-Net with $d$ encoder/decoder blocks, starting with $f$ filters in the first block.

| Name | Number of filters per encoder layer | Parameter Count | % Increase over U$N$-32 |
|---|---|---|---|
| U3-32 | 32,64,128 | 12,062,476 | - |
| U3-64 | 64,128,256 | 16,496,140 | 36.75 |
| U3-128 | 128,256,512 | 30,807,052 | 155.39 |
| U3-256 | 256,512,1024 | 81,203,212 | 573.18 |
| U4-32 | 32,64,128 | 16,556,044 | - |
| U4-64 | 64,128,256 | 31,053,836 | 87.56 |
| U4-128 | 128,256,512,1024 | 82,204,684 | 396.52 |
| U4-256 | 256,512,1024,2048 | 273,127,436 | 1549.71 |
| U5-32 | 32,64,128,256,512 | 31,113,740 | - |
| U5-64 | 64,128,256,512,1024 | 82,451,468 | 165.00 |
| U5-128 | 128,256,512,1024,2048 | 274,128,908 | 781.05 |
| U5-256 | 256,512,1024,2048,4096 | 1,013,491,724 | 3157.37 |
| U6-32 | 32,64,128,256,512,1024 | 82,511,372 | - |
| U6-64 | 64,128,256,512,1024,2048 | 274,375,692 | 232.5 |

In addition to the model parameters themselves, the total training time for a model is affected by the I/O pipeline and the hardware capabilities. Analysis of file I/O for training is not the focus of this paper. From a hardware perspective, the use of mixed precision compute for model training can significantly improve training time. We trained our implementations on NVIDIA Volta GPUs which can speed up linear and convolutional layers by using tensor cores on the GPU. While enabling a massive amount of compute, GPUs also tend to impose restrictions on model training primarily due to the limited amount of memory available on the device. For a given model and a given data size, the total amount of memory available on a GPU tends to limit the largest batch size that can used for training and this in turn can lead to increased training time for small batch sizes.

Taken together, the training time for a model is thus affected by design choices that may force trade-offs between training time and model accuracy. It is important to carefully weigh these trade-offs when selecting an optimal architecture.

## IV. EVALUATION METRICS

The section describes the set of quantitative evaluation metrics used for assessing the quality of the nowcasting model. The output of the nowcast model consists of 12 images representing the predicted weather for the next hour. The model is expected to produce textural detail as well as motion in the radar data. Given that image quality assessment can be inherently challenging, we use forecast-specific metrics as well as commonly used image quality metrics for evaluating networks trained on the SEVIR dataset. The overall quality of the generated `vil` imagery is evaluated using metrics common in forecast evaluation [12]. These metrics are computed by first binarizing the truth and prediction images at a set of thresholds that span the range 0 - 255. Thresholds for `vil` were chosen based on the 6 Video Integrator and Processor (VIP) intensity levels [13] which correspond to pixel values [16, 74, 133, 160, 181, 219]. Binarized pixels are scored as *Hits* (H) if `prediction=truth=1`, *Misses* (M) if `prediction=0,truth=1`, *False Alarms* (FA) if `prediction=1,truth=0` and *Correct Rejection* otherwise. The summary statistics calculated over the test set are : Probability of Detection (POD), Success Rate (SUCR), Critical Success Index (CSI), and BIAS. The statistics are computed as follows - $POD = \frac{H}{H+M}$, $SUCR = \frac{H}{H+FA}$, Critical Success Index (CSI)=$\frac{H}{H+M+FA}$, $BIAS = \frac{H+FA}{H+M}$

Evaluation of the perceptual quality of the output can be performed using several metrics such as Mean Squared Error (MSE) and Structural Similarity Index (SSIM) [14]. However, while pixel-level metrics are computationally efficient, they do not capture the rich textural information in weather radar images. Thus, to evaluate perceptual quality, we use the Learned Perceptual Image Path Similarity (LPIPS) metric proposed by Zhang et. al [15] which has been shown to outperform traditionally used metrics. The LPIPS evaluates images by computing the cosine distance between normalized network activations from deep networks such as AlexNet, SqueezeNet and VGG. In this paper we use an AlexNet network trained on the ImageNet [16] dataset. In the nowcast application, metrics were averaged over the 12 steps in the output. Metrics reported in the paper were calculated on an independent validation set consisting of 12,144 `vil` sequences.

## V. RESULTS

Experiments were performed on the Lincoln Laboratory Supercomputing Center (LLSC) TX-Gaia supercomputer. This cluster consists of 448 computer nodes with dual Intel Xeon Gold 6248 CPUs with 384 GB of RAM and two NVIDIA Volta V100 GPUs with 32 GB of memory per GPU. All models were trained using GPUs. Models were developed in TensorFlow 2.1 and the Horovod framework was used for data distributed training. All models were trained on 44,760 VIL image sequences generated from the SEVIR dataset. The input to the models consists of 13 images of size 384x384 pixels representing one hour of data. The model output consists of 12 images of the same pixel dimensions representing the forecast for the next hour. A separate test dataset consisting of 12,144 sequences was used to calculate the evaluation metrics discussed in Section IV. We train different configurations of the nowcast model as listed in Table I.

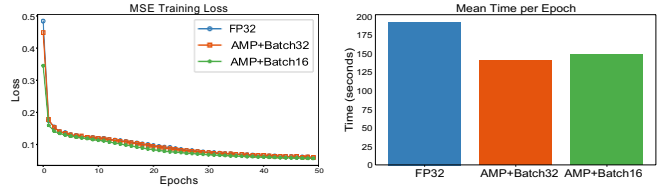

Fig. 2. Full vs. automatic mixed-precision training: Loss profile is seen to be unchanged when using mixed-precision training as compared with training in full 32-bit floating point precision. Using mixed-precision leads to a reduced mean epoch time over the training process as seen here.

### A. Mixed-precision training

In order to establish a baseline for compute performance, we trained the U4-32 U-Net implementation with 32-bit floating point precision as well as 16-bit mixed-precision. The mixed-precision model was trained with batch sizes of 32 and 16 whereas the full precision model was only trained with a batch size of 32. Using larger batch sizes results in out- of-memory (OOM) errors on the GPU. Figure 2 shows the loss curves during training as well as the average time per epoch for these three model configurations. It can be seen here that there the training loss is barely affected by switching to mixed-precision training. Moreover, mixed-precision training provides a 26.89% and 22.07% speedup for batch sizes of 32 and 16 respectively when compared with the original full precision trained model as shown in Figure 2. We do not report the total training time here because this also includes file I/O which we are not evaluating in this paper.

Table II shows GPU usage statistics obtained with the NVIDIA DGCM [17] tools across several U-Net architecture variations. One interesting observation here is that simply switching to mixed-precision training for the same model results not only in a reduction of the per epoch compute time, but it is achieved at a lower resource utilization as seen by the average Streaming Multiprocessor (SM) and GPU memory utilization in Table II. The lower numbers for SM utilization suggests that this particular configuration of the model is leading to the GPUs being data starved. If the SMs had more data available to compute or a larger number of compute operations to hide overheads, one could expect the SM utilization to increase as well. Increased compute operations can be achieved by either increasing the number of layers or the number of filters per layer or some combination of the two. In some cases, deeper and wider networks are able to perform better at a given task. Faster training through AMP enables wider parameter sweeps which can lead to the identification of improved model architectures which perform better (albeit at a higher compute cost).

### B. Comparison of U-Net model performance

While our original model shows modest reduction in compute time with mixed-precision training, the model suffers from the use of small encoder layers. Empirically, deeper and wider models have also been shown to achieve improved results as seen by their successes in the ImageNet [16] challenge. Thus, we can potentially improve nowcast performance while also leveraging compute optimizations by making our model larger. We evaluated other U-Net model configurations with increased encoder filter sizes as shown in Table I. These models are also trained using mixed-precision compute. The compute time speedups observed per epoch are listed in Table III and shown graphically in Figure 3 against the number of parameters in each model. Generally, wider models with larger number of filters achieved the biggest advantage with AMP. However, larger models also require a reduction in the batch size used which may increase compute time per epoch as compared with the smaller models.

TABLE II
GPU utilization statistics from NVIDIA DCGM tool: AMP refers to the use of Automatic Mixed Precision for training models.

| Model | Avg. SM Utilization % | | Max. Memory Utilization % | | Avg. Memory Utilization % | | Energy Usage (Joules) | |
|---|---|---|---|---|---|---|---|---|
| | FP32 | AMP | FP32 | AMP | FP32 | AMP | FP32 | AMP |
| U3-32 | 61.5 | 55 | 40.5 | 34 | 90.5 | 84.0 | 991,870.0 | 834,312.0 |
| U3-64 | 78.0 | 67 | 49 | 42 | 75.0 | 90.5 | 1,169,834.0 | 1,026,087.0 |
| U3-128 | 89.5 | 81.5 | 49.5 | 44 | 76.5 | 76.5 | 1,318,431.0 | 1,255,385.0 |
| U3-256 | 96.0 | 94 | 38 | 37 | 100 | 59.5 | 1,540,444.0 | 1,428,421.0 |
| U4-32 | 64.5 | 59 | 40 | 34 | 84 | 75.5 | 959,247.0 | 826,726.0 |
| U4-64 | 77 | 70.5 | 48 | 42.5 | 75 | 87 | 1,163,470.0 | 1,057,388.0 |
| U4-128 | 91 | 88.5 | 62 | 39 | 84.5 | 62.5 | 1,416,633.0 | 1,255,483.0 |
| U4-256 | 93 | 94.5 | 73.5 | 32 | 100 | 59.5 | 1,474,127.0 | 1,349,416.0 |
| U5-32 | 63.0 | 58.0 | 41.0 | 34.0 | 83.5 | 74.0 | 951,037.0 | 815,611.0 |
| U5-64 | 80.5 | 71.5 | 47.0 | 41.0 | 75.0 | 83.5 | 1,156,716.0 | 1,098,744.0 |
| U5-128 | 93.0 | 88.0 | 54.0 | 34.5 | 95.0 | 73.5 | 1,307,724.0 | 1,179,121.0 |
| U6-32 | 67.5 | 62.0 | 66.0 | 65.0 | 28.0 | 26.0 | 740,556.0 | 538,298.0 |
| U6-64 | 94.0 | 89.0 | 64.5 | 58 | 26.0 | 20.5 | 796,253.0 | 691,416.0 |

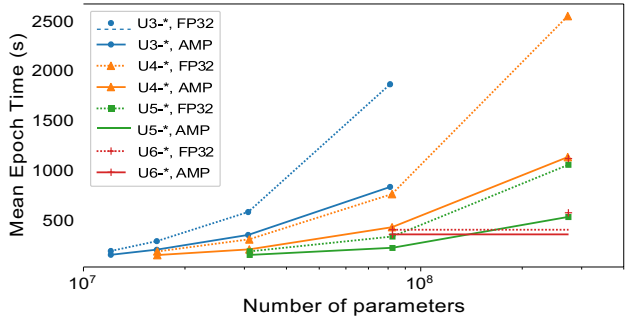

Fig. 3. Performance comparisons of FP32 vs AMP across different U-Net configurations. The "wider" networks (those with more filters in each layer) show much larger performance gains using AMP.

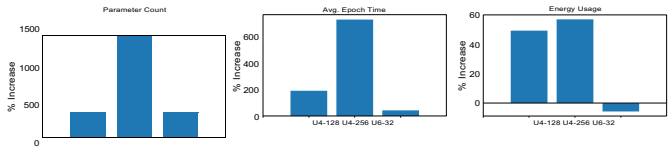

Fig. 4. Comparison of parameters, mean epoch time and energy usage for U4-128, U4-256, U6-32 model configurations relative to the U4-32 configuration. All comparisons in this figure used results from mixed-precision training for all the networks.

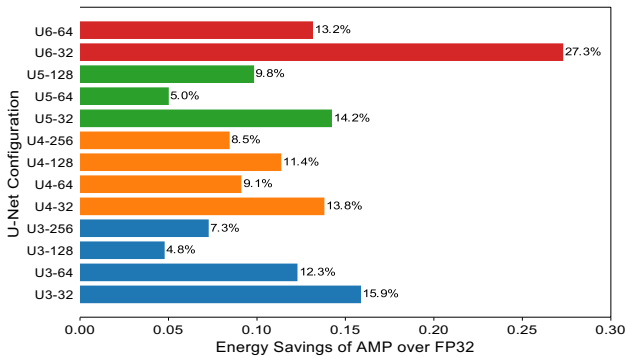

Fig. 5. Reductions in energy usage by training UNets with AMP

TABLE III
Speedups from mixed-precision training for different models. AMP: Automatic mixed-precision

| Model | Batch Sz. | Mean Epoch Time (Sec.) | | FP16 |
|---|---|---|---|---|
| | FP32/AMP | FP32 | AMP | Speedup % |
| U3-32 | 48 | 179.093 | 140.338 | 21.63 |
| U3-64 | 16/32 | 277.04 | 191.588 | 30.84 |
| U3-128 | 16 | 570.643 | 339.558 | 40.49 |
| U3-256 | 8 | 1853.6725 | 821.852 | 125.54 |
| U4-32 | 32 | 173.485 | 137.725 | 25.96 |
| U4-64 | 16/32 | 295.35 | 194.518 | 34.14 |
| U4-128 | 8/8 | 749.98 | 417.005 | 44.39 |
| U4-256 | 4/8 | 2537.88 | 1121.4 | 55.81 |
| U5-32 | 32/32 | 176.50 | 139.26 | 21.10 |
| U5-64 | 16/32 | 325.10 | 210.21 | 35.34 |
| U5-128 | 8/16 | 1045.185 | 520.845 | 100.67 |
| U6-32 | 4 | 392.87 | 345.86 | 13.59 |
| U6-64 | 4 | 1106.48 | 562.92 | 95.55 |

Table 2 also shows the GPU usage statistics for the larger

U-Net models. With the U4-128 and U4-256 configurations we observe that the average SM utilization on the GPU is significantly higher than the U4-32 model. Figure 4 shows the percentage increases in the mean epoch time and energy usage for U4-128, U4-256, U6-32 model configurations relative to the U4-32 configuration when all models were trained using mixed-precision. It can be seen that a nearly 400% increase in the parameter count for the U4-128 model results in a 200% increase in mean epoch time and only 45% increase in the energy usage.

An additional benefit of using AMP training is a modest reduction in the total energy used to train the model. The last two columns of Table II show the total energy usage in Joules measured when training each network in FP32 and AMP for a fixed number of epochs. The resulting reduction in energy usage in percent is plotted in Figure 5 for each configuration, which shows reductions ranging between 4.8 to 27.3%. While AMP appears to decrease energy usage when training with a fixed number of epochs, it is unclear if this reduction has any dependence on network configuration or size (in contrast to training time per epoch). The low energy usage observed in the U6 architectures was also surprising, and requires further investigation.

*C. Nowcast Evaluation*

As seen in the previous sub-section, using mixed-precision training can offer significant improvements in training time for larger models. However, there is a risk that training with lower precision might negatively impact the quality of the results. In order to test this potential trade-off, we evaluated the model performance for precipitation nowcasting for a subset of the U-Net architectures. Figure 6 shows example outputs from some of the models trained using mixed-precision compute. The top row in this figure shows the expected model output. Each successive row shows the output of different models. Columns correspond to the time step in the output and successive columns represent weather at 5 minute intervals. A visual inspection of these results shows that the larger models (U4-128 and U4-256) appear to perform better than the smaller U4-32 model. Note that larger U-Net configurations have significantly more trainable parameters and can potentially benefit from longer training runs.

Figure 7 shows that the time-savings achieved with AMP comes at minimal reduction in nowcast performance as measured with metrics from Section IV. These plots show evaluation on three U-Net configurations trained with AMP, relative to the baseline U4-32 model, as well as the persistence forecast (which is obtained by simply repeating the final image in the input sequence). The metrics plotted here include MSE, LPIPS, as well as POD, SUCR and CSI at two thresholds (74 and 133). All models trained outperformed the persistence forecast. Each models becomes worse as one goes forward in lead time, which is to be expected. Overall, the models trained with FP16 show no obvious disadvantage relative to FP32. The U6-32 model trained with FP16 AMP shows the best performance in the majority of metrics, demonstrating the value of using AMP training for this problem.

## VI. Conclusion

In this paper we discuss the implementation of U-Net based deep neural network for precipitation nowcasting and the design choices that influence the performance of the model. We show that using mixed-precision training for these models can provide a significant boost in compute performance with an appropriate selection of the model architecture. The combination of data distributed and mixed-precision training allows us to rapidly prototype a variety of models. We show that increasing model complexity by 1549.71% as represented by the number of trainable parameters in the U4-256 model only requires a 714.23% increase in training time as compared with the U4-32, model when using mixed-precision training. Perhaps more importantly, this is accompanied by only a 63.22% increase in energy used to train the model on the same dataset, for the same number of epoch without sacrificing model performance. Most surprisingly, energy usage actually decreases by appropriate selection of model configuration as seen in the case of U4-32 and U6-32 in Table II. While a comparison of loss functions and the effect on energy usage and training time was not the focus of this paper, this is part of our ongoing research. Finally, the experiments described in this paper focus on a meteorology application that uses variations of a fully-convolutional U-Net model architecture which is widely used in diverse application domains. We expect that these findings can help inform applications in other areas where large datasets and model configurations lead to challenges for design choices. Future work involves the investigation of other model architectures such as convolutional LSTMs and recurrent architectures for modeling the nowcast problem.

## VII. Acknowledgements

The authors wish to acknowledge the following individuals for their contributions and support:Bob Bond, Steve Rejto and Dave Martinez, along with William Arcand, David Bestor, William Bergeron, Chansup Byun, Matthew Hubbell, Vijay Gadepally, Michael Houle, Jeremy Kepner, Anna Klein, Peter Michaleas, Lauren Milechin, Julie Mullen, Andrew Prout, Albert Reuther, Antonio Rosa, and Charles Yee.

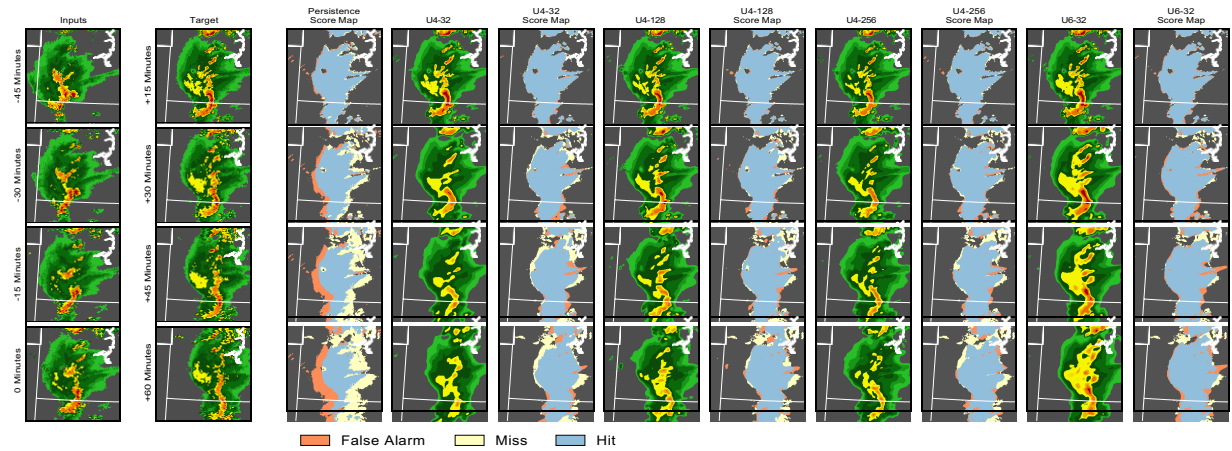

Fig. 6. Model outputs for different U-Net configurations: Each row in this figure shows the predicted output from the different models evaluated in this paper. Columns represent successive time steps. As seen here, larger models are able to produce greater amounts of texture in the predicted images as compared with the baseline U4-32 model. However, all models fail to produce sufficient detail in the last few time steps. Improving the outputs of the model farther into the future is the focus of future research.

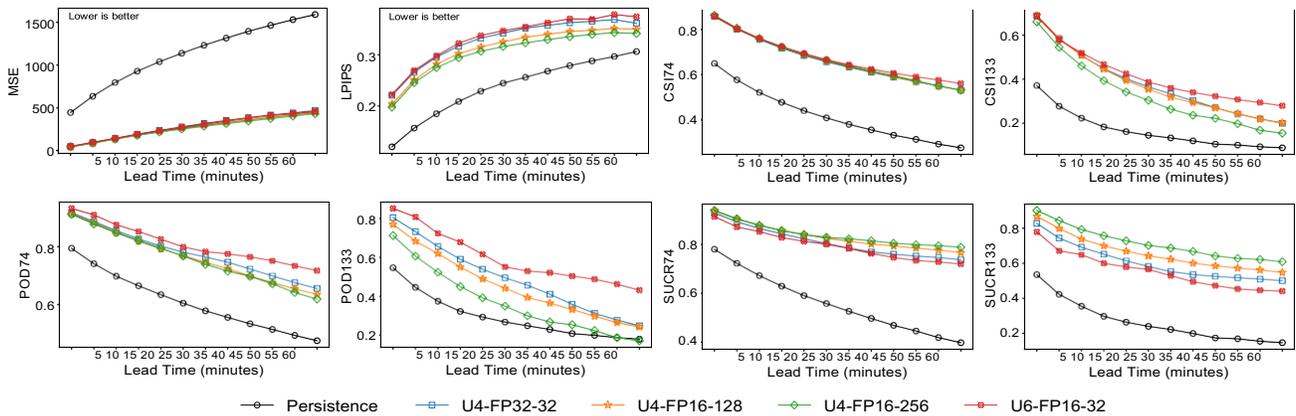

Fig. 7. Comparison of different models using domain specific and image quality metrics: The *Persistence* model represents a trivial model that simply repeats the last image seen. The LPIPS [18] metric evaluates the perceptual quality of the produced image. In this figure, FP16 refers to the use of Automatic Mixed Precision (AMP) for training. All other metrics are computed after binarization of the truth and predicted data as described in Section IV.